\documentclass[a4paper,twocolumn,11pt,accepted=2022-07-24]{quantumarticle}
\pdfoutput=1
\usepackage[utf8]{inputenc}
\usepackage[english]{babel}
\usepackage[T1]{fontenc}
\usepackage{amsmath}
\usepackage{hyperref}

\usepackage{tikz}
\usepackage{lipsum}

\usepackage{wrapfig}
\usepackage{float}
\usepackage{amssymb}
\usepackage{amsfonts}
\usepackage{amsmath}
\usepackage{cases}
\usepackage{stmaryrd}
\usepackage{bbold}
\usepackage{physics}
\usepackage{mathtools}

\begin{document}

\title{Coplanar Antenna Design for Microwave Entangled Signals Propagating in Open Air}

\author{Tasio Gonzalez-Raya}
\affiliation{Department of Physical Chemistry, University of the Basque Country UPV/EHU, Apartado 644, 48080 Bilbao, Spain}
\affiliation{EHU Quantum Center, University of the Basque Country UPV/EHU}
\orcid{0000-0001-9645-9681}
\author{Mikel Sanz}
\email{mikel.sanz@ehu.eus}
\affiliation{Department of Physical Chemistry, University of the Basque Country UPV/EHU, Apartado 644, 48080 Bilbao, Spain}
\affiliation{EHU Quantum Center, University of the Basque Country UPV/EHU}
\affiliation{Basque Center for Applied Mathematics (BCAM), Alameda de Mazarredo 14, 48009 Bilbao, Basque Country, Spain}
\affiliation{IKERBASQUE, Basque Foundation for Science, Plaza Euskadi 5, 48009 Bilbao, Spain}
\orcid{0000-0003-1615-9035}
\maketitle

\begin{abstract}
Open-air microwave quantum communication and metrology protocols must be able to transfer quantum resources from a cryostat, where they are created, to an environment dominated by thermal noise. Indeed, the states carrying such quantum resources are generated in a cryostat characterized by a temperature $T_\text{in} \simeq 50 $~mK and an intrinsic impedance $Z_\text{in} = 50 \, \Omega$. Then, an antenna-like device is required to transfer them with minimal losses into open air, characterized by an intrinsic impedance of $Z_\text{out} = 377 \, \Omega$ and a temperature $T_\text{out} \simeq 300$~K. This device accomplishes a smooth impedance matching between the cryostat and the open air. Here, we study the transmission of two-mode squeezed thermal states, developing a technique to design the optimal shape of a coplanar antenna to preserve the entanglement. Based on a numerical optimization procedure, we find the optimal shape of the impedance, and we propose a functional ansatz to qualitatively describe this shape. Additionally, this study reveals that the reflectivity of the antenna is very sensitive to this shape, so that small changes dramatically affect the outcoming entanglement, which could have been a limitation in previous experiments employing commercial antennae. This work is relevant in the fields of microwave quantum sensing and quantum metrology with special application to the development of the quantum radar, as well as any open-air microwave quantum communication protocol.
\end{abstract}

\section{Introduction}
Superconducting circuit technology, working in the microwave regime, has been around for a few decades but it is modern when compared with quantum optics. Recently, it has gained new life thanks to advances in controllability and scalability of superconducting qubits~\cite{Koch2007}, spurring this technology to the top on the field of quantum computation~\cite{Arute2019}. The development of quantum microwave technology is then vital not only for quantum computation, but also for secure quantum communication protocols~\cite{Fedorov2018, Pogorzalek2019, Fedorov2021,Fesquet2022,GonzalezRaya2022}, distributed quantum computing~\cite{Cuomo2020}, quantum metrology and quantum sensing~\cite{Sanz2017,LasHeras2017,Reichert2022}, specially with the quantum radar on sight~\cite{Sanz2018, Casariego2022}. 

The quantum microwave technology toolbox is constantly being updated, including new generations of HEMTs~\cite{Mariantoni2010,Eichler2011,Menzel2012}, JPAs~\cite{Pogorzalek2018,Fedorov2016}, and more recently, single-photon photodetectors~\cite{Sathyamoorthy2016,Kono2018} and photocounters~\cite{Dassonneville2020}.

Quantum communication with microwave photons is the best way to connect several superconducting-qubit chips together, as envisioned by the area of distributed quantum computing. Although the number of thermal photons is larger in the microwave than in the optical regime, attenuation of signals in the atmosphere is reduced highly in the frequency window 100 MHz - 10 GHz~\cite{Sanz2018,Lesurf1995}.

Applications of open-air quantum communication and sensing are particularly challenging and require additional development, specially for antennas connecting cryostats with the open air. Recent experiments have failed on efficient entanglement distribution while using commercial antennae~\cite{SandboChang2019,Luong2020,Barzanjeh2020}.

One could then wonder what are the limitations of using classical antennae transmitting quantum signals. A classical antenna comprises two mechanisms: impedance matching and amplification~\cite{Pozar2012}. While the former will be also required by quantum signals, in order to avoid reflections when propagating between different mediums, the latter can only destroy quantum correlations. Phase-insensitive amplification simply pollutes the signal with thermal noise~\cite{DiCandia2015}, thus diminishing quantum correlations. Therefore, a quantum antenna must not introduce this type of gain into the signal, and it needs to be specifically designed for the problem of entanglement preservation of signals traveling from inside a cryostat into the open air.

\begin{figure}[t]
{\includegraphics[width=0.45 \textwidth]{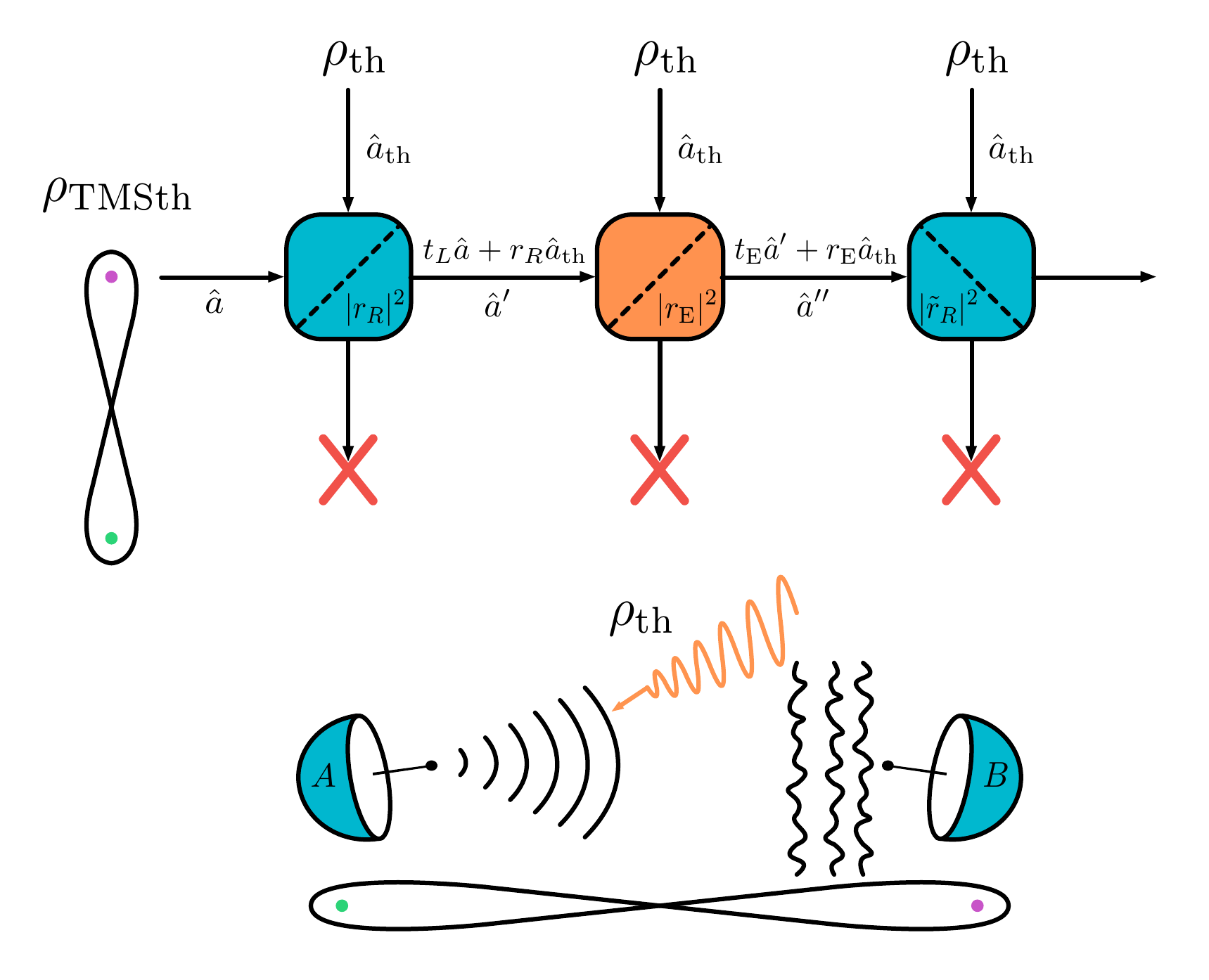}}
\caption{Sketch describing an open-air microwave quantum communication protocol, in which a party A generates a two-mode entangled quantum state and sends one of the modes to a second party, B, through an environment dominated by thermal noise, keeping the other mode. The effect of the antennae, as well as the transmission in open air, are modeled by beamsplitters, which allow for the description of the deterioration of the state due to thermal noise. These have reflectivities $|r_{R}|^{2}$ and $|\tilde{r}_{R}|^{2}$ for the antennae, and $|r_{E}|^{2}$ for the imperfect open-air transmission.}
\label{fig1}
\end{figure}

In this article, we address this problem by considering the quantum antenna as a coplanar waveguide with a position-dependent impedance. We observe that the shape of the antenna defines its reflectivity, and this highly affects entanglement. Thus, we optimize the impedance function with the objective of minimizing the reflectivity of the antenna. As a paradigmatic case, we study the transmission of two-mode squeezed states~\cite{Kim2002} into open air, for they are easy to generate and robust to photon losses. We employ a numerical optimization method through interpolation and repurposing each solution, as well as propose an ansatz for the impedance, qualitatively-based on the solution from the numerical case. We find that the reflectivity can be reduced below $10^{-9}$, while entanglement preservation with real-life experimental parameters would require values below $10^{-4}$. To conclude, we study the dependence of entanglement preservation on errors in the impedance of the antenna, to illustrate the impact that small fabrication imperfections could have on the quantum antenna.

\section{Antenna design}
\begin{figure*}[t]
\centering
{\includegraphics[width=0.85 \textwidth]{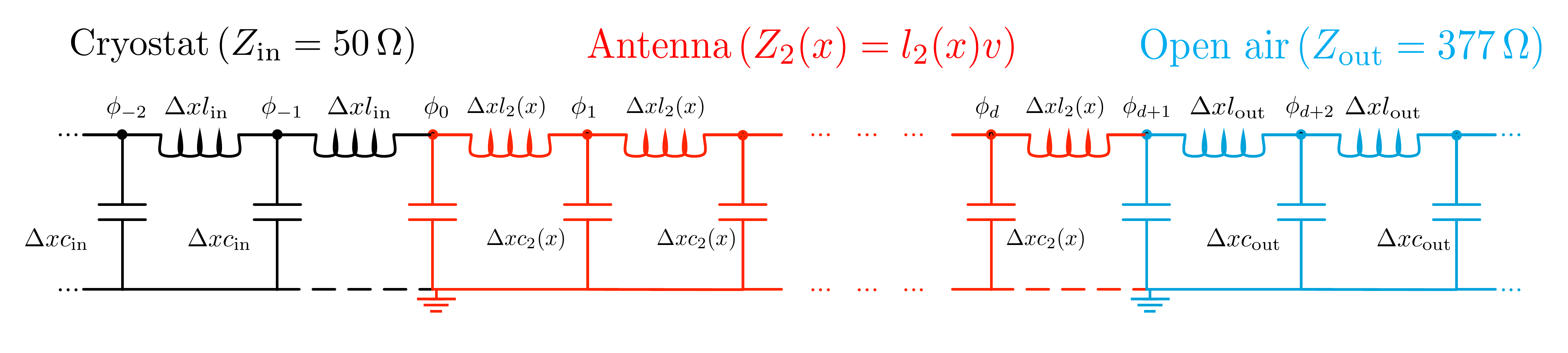}}
\caption{Description of a quantum circuit that represents the connection of a cryostat (black) with the open air (blue), represented by a transmission waveguide, via the antenna (red), which is described by a finite inhomogeneous transmission line.}
\label{fig2}
\end{figure*}

We intend to design an antenna for an open-air microwave quantum communication protocol, in which an entangled state is produced by a source A, keeping one mode and sending another through a waveguide into open air, to be received at a remote location B, while maintaining the entanglement between both modes, as can be seen in Fig.~\ref{fig1}. For this, we propose a transmission line (TL) as a waveguide that sends out the state, then a finite inhomogeneous transmission line as the antenna, and then another TL to represent the open air~\cite{Sanz2018}. This circuit is sketched in Fig.~\ref{fig2}.

The TL on the left has an impedance of $50\,\Omega$, whereas that on the right has an impedance of $377\,\Omega$. Then, the antenna serves as an inhomogeneous medium that achieves a smooth transition from two very different impedances. The Lagrangian describing this circuit is
\begin{widetext}
\small
\begin{equation*}
\mathcal{L} = \sum_{i=-N}^{-1} \left[ \frac{\Delta x \, c_{\text{in}}}{2}\dot{\phi}_{i}^{2} - \frac{(\phi_{i+1}-\phi_{i})^{2}}{2\Delta x \, l_{\text{in}}} \right] +  \sum_{j=0}^{d} \left[ \frac{\Delta x \, c_{2}(x)}{2}\dot{\phi}_{j}^{2} - \frac{(\phi_{j+1}-\phi_{j})^{2}}{2\Delta x \, l_{2}(x)} \right] + \sum_{k=d+1}^{N} \left[ \frac{\Delta x \, c_{\text{out}}}{2}\dot{\phi}_{k}^{2} - \frac{(\phi_{k+1}-\phi_{k})^{2}}{2\Delta x \, l_{\text{out}}} \right],
\end{equation*}
\end{widetext}
where we have defined $l_{\text{in}}, c_{\text{in}}$ as the inductance and capacitance densities of the transmission line inside the cryostat, $l_{2}(x), c_{2}(x)$ as the inductance and capacitance densities of the antenna, and $l_{\text{out}}, c_{\text{out}}$ as the inductance and capacitance densities of the second transmission line. See that the inductances and capacitances on the antenna depend on the position, which is necessary for a smooth change of impedance. 
This way, we can write the impedance as
\begin{equation}
Z_{2}(x) = \sqrt{\frac{l_{2}(x)}{c_{2}(x)}} = l_{2}(x) v.
\end{equation}
The reflectivity of the antenna entirely depends on the impedance between the different media. As the impedance is defined by the densities of inductance and capacitance, $Z = \sqrt{l/c}$, which can be independently manipulated in nano fabrication, we can choose without a loss of generality the propagation velocity through the antenna, $v=1/\sqrt{lc}$, to be constant. Consequently, the dependence on the position falls entirely onto the impedance.

Taking $N\rightarrow\infty$ in order to consider semi-infinite transmission lines, amounts to taking the continuum limit $\Delta x\rightarrow 0$. Then, we rewrite the Lagrangian
\begin{equation}
\mathcal{L} = \int_{-\infty}^{\infty} dx \left[ \frac{c(x)}{2} (\partial_{t}\phi(x,t))^{2} - \frac{1}{2l(x)} (\partial_{x}\phi(x,t))^{2} \right],
\end{equation}
defining the capacitances and inductances as
\begin{equation}
l(x) = \begin{cases} 
l_{\text{in}} & \text{if } x < 0 \\ 
l_{2}(x) & \text{if } 0 \leq x \leq d \\ 
l_{\text{out}} & \text{if } x > d 
\end{cases}
\end{equation}
and 
\begin{equation}
c(x) = \begin{cases} 
c_{\text{in}} & \text{if } x < 0 \\ 
c_{2}(x) & \text{if } 0 \leq x \leq d \\ 
c_{\text{out}} & \text{if } x > d 
\end{cases}
\end{equation}
From the minimal action principle, we obtain the Euler-Lagrange equations for this Lagrangian, 
\begin{equation}\label{E-L}
c(x) \partial_{t}^{2}\phi(x,t) = \partial_{x}\left(\frac{\partial_{x}\phi(x,t)}{l(x)}\right).
\end{equation}
For the left and right transmission lines, $l(x)$ and $c(x)$ are constant, and Eq.~\ref{E-L} is just the wave equation. This means that for the left and right TLs, the solutions to the equations of motion are plane waves. However, the solution for the antenna is not as straightforward, for which we employ the variable separation method. We then propose $\phi(x,t)=\sum_{n}\varphi_{n}(t)u_{n}(x)$, 
\begin{equation}
c(x) \ddot{\varphi}_{n}(t)u_{n}(x) = \varphi_{n}(t)\left( \frac{l(x)u''_{n}(x)-l'(x)u'_{n}(x)}{l^{2}(x)} \right),
\end{equation}
which leads to the expression
\begin{equation}
c(x) l(x) \frac{\ddot{\varphi}_{n}(t)}{\varphi_{n}(t)} = \frac{1}{u_{n}(x)}\left( u''_{n}(x) - \frac{l'(x)}{l(x)} u'_{n}(x) \right).
\end{equation}
On both sides of the equation, the solutions are constants, 
\begin{eqnarray}
&& \ddot{\varphi}_{n}(t) = -\xi v^{2} \varphi_{n}(t), \\
&& u''_{n}(x) - \frac{Z'(x)}{Z(x)} u'_{n}(x) = -\xi u_{n}(x)
\end{eqnarray}
where we have used $c(x)l(x)=1/v^{2}$ and $Z(x)=l(x)v$. From this, we see that $\xi=k_{n}^{2}=(\omega_{n}/v)^{2}$, the wavenumber. Then, the equation that we need to solve is that for $u_{n}(x)$, which can be written as the Sturm-Liouville problem. In order to solve this equation for the antenna we need to fix $Z(x)$. 

\subsection{Linear antenna}
For a simple case study we consider that $Z(x)$ for $0\leq x\leq d$ is a linear function of the position,
\begin{equation}
Z(x) = \left(1-\frac{x}{d}\right)Z_{\text{in}} + \frac{x}{d} Z_{\text{out}},
\end{equation}
which implies that the inductance in the antenna is also linear. From now on, we will focus on a single mode of the wavefunction, and we will drop the subscript notation $\cdot_{n}$. The solution to the equation of motion (see Appendix~\ref{app1}) is 
\begin{eqnarray}\label{SL_solution}
\nonumber u(x) &=& \left( Z_{\text{in}}(d-x) + Z_{\text{out}}x\right) \times \\
\nonumber && \bigg[ c_{1} J_{1}\left(kx + kd\frac{Z_\text{in}}{Z_\text{out}-Z_\text{in}}\right) \\
&+& c_{2} Y_{1}\left(kx + kd\frac{Z_\text{in}}{Z_\text{out}-Z_\text{in}}\right) \bigg],
\end{eqnarray}
where $J_{1}(\cdot)$, $Y_{1}(\cdot)$ are the Bessel functions of the first and second kind, respectively, and $c_{1}$, $c_{2}$ are arbitrary constants. Then, our problem can be translated into a scattering problem, 
\begin{eqnarray}
\nonumber && u_{1}(x) = A e^{ikx} + B e^{-ikx}  \qquad \text{for } x<0\\
\nonumber && u_{2}(x) = u(x) \qquad \text{for } 0 \leq x \leq d\\
&& u_{3}(x) = F e^{iqx} + G e^{-iqx}  \qquad \text{for } x > d
\end{eqnarray}
where $k=\omega/(c/3)$ is the wavenumber inside the cryostat and the antenna, considering that the propagation velocity is $v=c/3$ inside these two circuits, and $q=\omega/c$ is the wavenumber on open air, where $v = c$. Imposing continuity of voltage
\begin{equation}
\left.\dot{\phi}(x,t)\right|_{x^{-}} = \left.\dot{\phi}(x,t)\right|_{x^{+}}
\end{equation}
and current
\begin{equation}
\left. \partial_{x}\left(\frac{\phi(x,t)}{l(x)}\right)\right|_{x^{-}} =\left. \partial_{x}\left(\frac{\phi(x,t)}{l(x)}\right)\right|_{x^{+}}
\end{equation}
implies imposing the continuity of these functions and their derivatives. Boundary conditions are imposed at frontier points $x=0$ and $x=d$, such that 
\begin{eqnarray*}
\nonumber \lim_{x\rightarrow0^{-}}u(x) &=& u_{1}(0), \qquad \lim_{x\rightarrow d^{-}}u(x) = u_{2}(d), \\
\lim_{x\rightarrow0^{+}}u(x) &=& u_{2}(0), \qquad \lim_{x\rightarrow d^{+}}u(x) = u_{3}(d),
\end{eqnarray*}
considering that the impedance is constant across the boundaries, $\lim_{x\rightarrow0^{-}}Z(x) = \lim_{x\rightarrow0^{+}}Z(x) = Z_{\text{in}}$ and $\lim_{x\rightarrow d^{-}}Z(x) = \lim_{x\rightarrow d^{+}}Z(x) = Z_{\text{out}}$. Notice that, since we have imposed that the velocity is constant throughout the antenna, it will jump from $v_{\text{in}} = c/3$ to $v_{\text{out}} = c$ when moving from the antenna to open air. Therefore, the continuity of the current at $x=d$ will be expressed by
\begin{equation*}
v_{\text{in}}\left. \partial_{x}\left(\frac{u_{2}(x)}{Z(x)}\right)\right|_{x=d^{-}} =\frac{v_{\text{out}}}{Z_{\text{out}}}\left. \partial_{x}u_{3}(x)\right|_{x=d^{+}}.
\end{equation*}
If we do this, we find 
\begin{eqnarray}
\nonumber A+B &=& d Z_{\text{in}} \bigg[ c_{1} J_{1}\left(kd\frac{Z_{\text{in}}}{Z_{\text{out}}-Z_{\text{in}}}\right) \\
&& + c_{2} Y_{1}\left(kd\frac{Z_{\text{in}}}{Z_{\text{out}}-Z_{\text{in}}}\right) \bigg], \\
\nonumber A - B &=& -i d Z_{\text{in}} \bigg[ c_{1} J'_{1}\left(kd\frac{Z_{\text{in}}}{Z_{\text{out}}-Z_{\text{in}}}\right) \\
&& + c_{2} Y'_{1}\left(kd\frac{Z_{\text{in}}}{Z_{\text{out}}-Z_{\text{in}}}\right) \bigg],
\end{eqnarray}
and also
\begin{eqnarray}
\nonumber F e^{iqd}+Ge^{-iqd} &=& d Z_{\text{out}} \bigg[ c_{1} J_{1}\left(kd\frac{Z_{\text{out}}}{Z_{\text{out}}-Z_{\text{in}}}\right) \\
 && +c_{2} Y_{1}\left(kd\frac{Z_{\text{out}}}{Z_{\text{out}}-Z_{\text{in}}}\right) \bigg], \\
\nonumber F e^{iqd}-Ge^{-iqd} &=& -i d Z_{\text{out}} \bigg[ c_{1} J'_{1}\left(kd\frac{Z_{\text{out}}}{Z_{\text{out}}-Z_{\text{in}}}\right) \\
&&  + c_{2} Y'_{1}\left(kd\frac{Z_{\text{out}}}{Z_{\text{out}}-Z_{\text{in}}}\right) \bigg],
\end{eqnarray}
where it will be useful to know that
\begin{equation}
J_{1}(x)Y'_{1}(x) - Y_{1}(x)J'_{1}(x) = \frac{2}{\pi x}.
\end{equation}
The transfer matrix
\begin{equation}
\begin{pmatrix} F \\ G \end{pmatrix} = T  \begin{pmatrix} A \\ B \end{pmatrix} 
\end{equation}
is used to construct a scattering matrix $S$,
\begin{equation}
{\small \begin{pmatrix} F \\ B \end{pmatrix}_{\text{out}} = S  \begin{pmatrix} A \\ G \end{pmatrix}_{\text{in}}  = \begin{pmatrix} S_{11} & S_{12} \\ S_{21} & S_{22} \end{pmatrix} \begin{pmatrix} A \\ G \end{pmatrix}_{\text{in}},}  
\end{equation}
which will not be normalized ($S S^{\dagger} \neq \mathbb{1}$). For that, we can redefine $S$ as
\begin{equation*}
{\small\bar{S} = \begin{pmatrix} \theta_{1} & 0 \\ 0 & \theta_{2} \end{pmatrix} S \begin{pmatrix} \theta_{1} & 0 \\ 0 & \theta_{2} \end{pmatrix} = \begin{pmatrix} \theta_{1}^{2} S_{11} & \theta_{1}\theta_{2} S_{12} \\ \theta_{1}\theta_{2} S_{21} & \theta_{2}^{2} S_{22} \end{pmatrix},}
\end{equation*}
and find the parameters $\theta_{1}$, $\theta_{2}$, with which the matrix $\bar{S}$ satisfies unitarity conditions. First of all, the determinant must be equal to one (in modulus). This implies that
\begin{equation}
\text{det}\,\bar{S} =  \theta_{1}^{2}\theta_{2}^{2} \text{det}\,S = e^{i\gamma}.
\end{equation}
Also, the rows of the matrix must represent orthonormal vectors,
\begin{eqnarray}
\nonumber && \begin{pmatrix} \theta_{1}^{2} S_{11} & \theta_{1}\theta_{2} S_{12} \end{pmatrix} \begin{pmatrix} \theta_{1}\theta_{2} S_{21}^{*} \\ \theta_{2}^{2} S_{22}^{*} \end{pmatrix} =\\
&& \theta_{1}\theta_{2}\left[ \theta_{1}^{2}S_{11}S_{21}^{*} + \theta_{2}^{2}S_{12}S_{22}^{*} \right],
\end{eqnarray} 
and from these two conditions we can obtain the parameters
\begin{eqnarray}
&& \theta_{1}^{4} = -\frac{e^{i\gamma}}{\text{det}\,S} \, \frac{S_{12}S_{22}^{*}}{S_{11}S_{21}^{*}}, \\
&& \theta_{2}^{4} = -\frac{e^{i\gamma}}{\text{det}\,S} \, \frac{S_{11}S_{21}^{*}}{S_{12}S_{22}^{*}}.
\end{eqnarray}
For this scattering problem, we find that the unitary scattering matrix is given by
\begin{equation}
{\small\bar{S} = \frac{e^{i\gamma/2}}{\sqrt{\text{det}\,S}}\begin{pmatrix} -\sqrt{\frac{ q Z_{\text{in}}}{ k Z_{\text{out}}}} S_{11} & S_{12} \\ S_{21} & -\sqrt{\frac{ k Z_{\text{out}}}{ q Z_{\text{in}}}} S_{22} \end{pmatrix},}
\end{equation}
where $\gamma$ is a free parameter of the system, and thus can be set to zero. Also we have that, for this problem, $|\text{det}\,S|=1$. In the entries of this matrix we can identify the transmission and reflection coefficients,
\begin{equation}
\bar{S} = \begin{pmatrix} t_{L} & r_{R} \\ r_{L} & t_{R} \end{pmatrix}.
\end{equation}
In the case in which there is no antenna, we now recover the usual formulas in optics
\begin{eqnarray}
&& |t_{L}|^{2} = |t_{R}|^{2} \rightarrow \left(\frac{2\sqrt{Z_{\text{in}}Z_{\text{out}}}}{Z_{\text{in}}+Z_{\text{out}}}\right)^{2}, \\
&& |r_{R}|^{2} = |r_{L}|^{2} \rightarrow \left(\frac{Z_{\text{in}}-Z_{\text{out}}}{Z_{\text{in}}+Z_{\text{out}}}\right)^{2},
\end{eqnarray}
when there is an abrupt change in the impedance. Notice that, when both impedances are equal, there are no reflections. In the opposite limit, having an infinitely-long antenna, we find that
\begin{eqnarray}
&& |t_{L}|^{2} = |t_{R}|^{2} \rightarrow 1, \\
&& |r_{R}|^{2} = |r_{L}|^{2} \rightarrow 0.
\end{eqnarray}
This limit corresponds to an infinitesimally-slow (adiabatic) change of impedance, generating no reflections in a wave propagating through it into another medium. Now, we want to obtain the transmission coefficient for a given state, depending on the size of the antenna, $d$. 

\section{Two-mode squeezed thermal states}
We study the performance of the antenna for two-mode squeezed states, the best candidate for entangled quantum states with continues variables due to the stability and simplicity with which they are generated. The entanglement of these states is determined by its squeezing parameter. A two-mode squeezed vacuum state results from the combination in a $50:50$ beam splitter of two single-mode squeezed vacuum states in different directions, generated with independent Josephson parametric amplifiers (JPAs). The action of JPA can be described by the single-mode squeezing operator,
\begin{equation}
S(z) = e^{\frac{1}{2}(z^{*}a^{2}-z a^{\dagger 2})},
\end{equation}
with $z=r e^{i\varphi}$. The single-mode squeezed vacuum states, with equal squeezing in orthogonal directions in phase space, are then sent through a $50:50$ beam splitter to obtain a two-mode squeezed vacuum state~\cite{Kim2002}. However, the presence of thermal photons during the generation of squeezed states in the JPAs, at temperatures of $10-50$ mK, leads to states which are not pure. To account for the presence of thermal photons, we consider that the input states to the JPAs are identical thermal states, with $n$ thermal photons, which leads to a two-mode squeezed thermal state (TMSth)~\cite{Serafini2004} through the same process.

In this setting, we will work with the covariance matrix, an object that contains the second moments of the state, since it is a good way to characterize gaussian states. This matrix can be written as
\begin{equation}
\sigma = \begin{pmatrix} \sigma_{1} & \sigma_{3} \\ \sigma_{3}^{T} & \sigma_{2} \end{pmatrix}
\end{equation}
by defining
\begin{eqnarray}
\nonumber \sigma_{1} &=&  \begin{pmatrix} \langle x_{1}^{2} \rangle & \frac{1}{2}\langle \{x_{1},p_{1}\} \rangle \\ \frac{1}{2}\langle \{x_{1},p_{1}\} \rangle &  \langle p_{1}^{2} \rangle \end{pmatrix}, \\
 \sigma_{2} &=&  \begin{pmatrix} \langle x_{2}^{2} \rangle & \frac{1}{2}\langle \{x_{2},p_{2}\} \rangle \\ \frac{1}{2}\langle \{x_{2},p_{2}\} \rangle &  \langle p_{2}^{2} \rangle \end{pmatrix}, \\
\nonumber\sigma_{3} &=& \frac{1}{2} \begin{pmatrix} \langle \{x_{1},x_{2}\} \rangle & \langle \{x_{1},p_{2}\} \rangle \\ \langle \{p_{1},x_{2}\} \rangle &  \langle \{p_{1},p_{2}\} \rangle\end{pmatrix}.
\end{eqnarray}
Consider a TMSth state generated by applying a two-mode squeezing operator, 
\begin{equation}
S_{12}(r) = \exp[r\left(a_{1}a_{2}-a_{1}^{\dagger}a_{2}^{\dagger}\right)],
\end{equation}
onto a thermal state. The covariance matrix of the resulting state is given by
\begin{equation}
\frac{1}{\sqrt{\mu}}\begin{pmatrix} c& 0 & s & 0 \\ 0 & c & 0 & -s \\  s & 0 & c & 0 \\ 0 & -s & 0 & c \end{pmatrix},
\end{equation}
where $c = \cosh 2r$, $s = \sinh 2r$, and $\mu$ is the purity, such that $\mu = 1/(1+2n)^{2}$ for TMSth states with $n$ thermal photons. Then, the covariance matrix of our initial state is
\begin{equation}
\sigma_{\text{in}} = \frac{1}{\sqrt{\mu}} \begin{pmatrix} c & 0 & s & 0 \\ 0 & c & 0 & -s\\  s & 0 & c & 0 \\ 0 & -s & 0 & c \end{pmatrix}.
\end{equation}
On the other hand, the covariance matrix of the thermal noise coming from the environment, which changes the state when it is sent into open air, is given by
\begin{equation}
\sigma_{\text{env}} = \begin{pmatrix}1+2N_{\text{eff}} & 0 \\ 0 & 1+2N_{\text{eff}} \end{pmatrix}.
\end{equation}
Here, $N_{\text{eff}}$ represents the number of thermal photons from the environment, that characterize the environmental noise in open air. The global covariance matrix is given by
\begin{widetext}
\begin{equation}
\sigma_{\text{env-in}} = (1+2n) \begin{pmatrix} \eta & 0 & 0 & 0 & 0 & 0 \\ 0 & \eta  & 0 & 0 & 0 & 0 \\ 0 & 0 &  \cosh 2r & 0 & \sinh 2r & 0 \\ 0 & 0 & 0 & \cosh 2r & 0 & -\sinh 2r \\ 0 & 0 & \sinh 2r & 0 & \cosh 2r & 0 \\ 0 & 0 & 0 & -\sinh 2r & 0 & \cosh 2r \end{pmatrix},
\end{equation}
\end{widetext}
where we have defined $\eta = (1+2N_{\text{eff}})/(1+2n)$. Now, in this quantum communication protocol we intend to send one of the modes of the TMSth state we generate, and optimize the entanglement remaining between the mode we have kept and the one that is sent into open air. Then, one of the modes goes through the antenna and is mixed with the thermal noise coming from the environment through the scattering matrix of the antenna, while the other mode is left untouched. This process is described by the action of the operator
\begin{equation}
T = \begin{pmatrix} B & 0 \\ 0 & \mathbb{1}_{2} \end{pmatrix}
\end{equation}
on the matrix above. See that thermal noise enters here as the second input mode to the beam splitter. We then trace out the reflected part coming from the beam splitter matrix, and obtain the covariance matrix of the output state,
\begin{equation}
\sigma_{\text{out}} = tr_{2} \left[ T\sigma_{\text{env-in}} T^{\dagger}\right].
\end{equation}
Given the order in which we have written the states in the covariance matrix, the beam splitter matrix $B$ is a reshuffling of the scattering matrix describing the action of the antenna, and can be written as
\begin{equation}
B =  \begin{pmatrix} r_{R}\mathbb{1}_{2} & t_{L}\mathbb{1}_{2} \\ t_{R}\mathbb{1}_{2} & r_{L}\mathbb{1}_{2} \end{pmatrix}.
\end{equation}
This way, we find the covariance matrix of the output state to be
\begin{widetext} 
\begin{equation}\label{CMout}
\sigma_{\text{out}} = (1+2n) \begin{pmatrix} \eta |r_{R}|^{2} + |t_{L}|^{2} \cosh 2r & 0 & t_{L} \sinh 2r & 0 \\ 0 & \eta |r_{R}|^{2} + |t_{L}|^{2} \cosh 2r & 0 & - t_{L} \sinh 2r \\ t_{L}^{*} \sinh 2r & 0 & \cosh 2r & 0 \\ 0 &  - t_{L}^{*} \sinh 2r & 0 & \cosh 2r \end{pmatrix}.
\end{equation}
\end{widetext}
In the covariance matrix formalism, we can compute the entanglement of a state through the negativity. For a covariance matrix
\begin{equation}
\sigma = \begin{pmatrix} \sigma_{1} & \sigma_{3} \\ \sigma_{3}^{T} & \sigma_{2} \end{pmatrix}
\end{equation}
the negativity is computed as
\begin{equation}
\mathcal{N} = \max\left[ 0, \frac{1-\nu}{2\nu}\right],
\end{equation}
where $\nu$ is the symplectic eigenvalue of the partial transposition of $\sigma$~\cite{Adesso2005}, 
\begin{equation}
\nu = \frac{1}{\sqrt{2}} \sqrt{\Delta(\sigma)-\sqrt{\Delta^{2}(\sigma)-4\det(\sigma)}},
\end{equation}
with $\Delta(\sigma) = \det(\sigma_{1}) + \det(\sigma_{2}) - 2\det(\sigma_{3})$. The state described by $\sigma$ is entangled if $\mathcal{N} > 0$, or equivalently $0 < \nu <1$. The symplectic eigenvalue of the TMSth state $\sigma_{\text{in}}$ is given by
\begin{equation}
\nu_{\text{in}} = (1+2n)e^{-2r},
\end{equation}
and the condition for entanglement is $r > \frac{1}{2}\log(1+2n)$. Notice that this condition is $r>0$ for two-mode squeezed vacuum states ($n=0$). 

The outgoing state will also be a TMSth state, up to unitary transformations. Thus, it's symplectic eigenvalue will have the same form as that of the initial state. If we set the initial squeezing to zero, no squeezing can be generated through the beam splitter, and we have $\nu_{\text{in}} = 1+2n$ for the initial state, and $\nu_{\text{out}} = 1+2n'$ for the final state. Comparing this formula with the symplectic eigenvalue obtained from Eq.~\ref{CMout}, we find that $n'=n$. Then, $\nu_{\text{out}} = (1+2n)e^{-2r'}$, and
\begin{equation}
r' = -\frac{1}{2}\log\left( \frac{\nu_{\text{out}}}{1+2n}\right).
\end{equation}
If we write out explicitly the symplectic eigenvalue of the output state, we obtain
\begin{equation*}
\nu_{\text{out}} = \frac{1}{\sqrt{2}} \sqrt{\Delta(\sigma_{\text{out}})- \sqrt{\Delta^{2}(\sigma_{\text{out}})-4\det(\sigma_{\text{out}})}}
\end{equation*}
where
\begin{eqnarray}
\nonumber \Delta(\sigma_{\text{out}}) &=& (1+2n)^{2} \Big[ (\eta|r_{R}|^{2}+|t_{L}|^{2}\cosh2r)^{2} \\
&& + \cosh^{2}2r + 2|t_{L}|^{2}\sinh^{2}2r \Big],
\end{eqnarray}
and $\sqrt{\Delta^{2}(\sigma_{\text{out}})-4\det(\sigma_{\text{out}})}$ is
\begin{eqnarray}
\nonumber &&  (1+2n)^{2} \Big[ \eta |r_{R}|^{2}+(1+|t_{L}|^{2})\cosh 2r \Big] \times \\
&& \sqrt{(\eta-\cosh2r)^{2}|r_{R}|^{4} + 4|t_{L}|^{2}\sinh^{2}2r}.
\end{eqnarray}
The number of thermal photons is computed from the Bose-Einstein distribution, 
\begin{equation}
n(f) \propto \frac{1}{e^{\frac{\hbar f}{k_{B}T}}-1}.
\end{equation}
The number of thermal photons of frequency $f = \omega/2\pi = 5$ GHz is $8\cdot10^{-3}$ at temperatures of $T\sim50$ mK, whereas at room temperature ($T\sim300$ K), the number of thermal photons is approximately $1250$, which implies $\eta\sim 2500$. Now, we can approximate $\nu_{\text{out}}$ depending on the relation between $\eta|r_{R}|^{2}$ and $|t_{L}|^{2}$, and obtain a simplified form in the different regimes. 

The first case we study describes the regime in which $\eta|r_{R}|^{2} \gg 1$ with $|r_{R}|\ne0$. We find that
\begin{eqnarray}
\nonumber \nu_{\text{out}} &=& \nu_{\text{in}} + (1+2n)\sinh2r \times \\
&& \left[ 1 - \frac{|t_{L}|^{2}(1+|t_{L}|^{2})\sinh^{2}2r}{2\eta|r_{R}|^{2}\cosh2r} \right].
\end{eqnarray}
Total reflection by the antenna is achieved by taking $|r_{R}|\rightarrow 1$, then $\nu_{\text{out}} = (1+2n)\cosh2r$, which is always greater or equal to 1. This means that there cannot be entanglement, because we are neglecting the reflected mode, and the transmitted one only has thermal noise from the environment. Then, we just have two thermal states. Total transmission, $|r_{R}|\rightarrow 0$, breaks the approximation we have made here.  Furthermore, see that we recover the result $1+2n$ as $r\rightarrow 0$. In this case, $\nu_{\text{out}}$ is smaller for larger $|t_{L}|$, only showing entanglement for $|r_{R}|<0.1$. This is the regime we study in the next case. 

The second case describes the scenario in which $\eta|r_{R}|^{2} \ll 1$ with $|t_{L}|\approx 1$. This regime is more restrictive, and is close to total transmission, simply because in order to have $\eta|r_{R}|^{2} \ll 1$ we need $|r_{R}| < 10^{-2}$. Here, we find
\begin{eqnarray}
\nu_{\text{out}} &=& \nu_{\text{in}}\left[ 1 + \frac{\eta|r_{R}|^{2}}{2}e^{2r} \right] \nonumber\\
&=&  \nu_{\text{in}} + \left( \frac{1}{2}+N_{\text{eff}}\right)|r_{R}|^{2}.
\end{eqnarray}
When $r\rightarrow 0$, the approximation breaks down and we would have to substitute before performing the approximation. For total transmission, $|r_{R}|=0$, we recover the initial state, since no thermal noise from the environment is mixed with the mode we are sending through the antenna. The condition for entanglement on the initial state is
\begin{equation}
r > \frac{1}{2}\log(1+2n),
\end{equation}
and in this case, for the output state it is
\begin{equation*}
r > \frac{1}{2}\log(1+2n) -\frac{1}{2}\log\left[ 1-  \left( \frac{1}{2}+N_{\text{eff}}\right)|r_{R}|^{2}\right],
\end{equation*}
which is, of course, more restrictive. The former inequality imposes $\left( \frac{1}{2}+N_{\text{eff}}\right)|r_{R}|^{2} < 1$. 

If we approximate $\log(1\pm x) = \pm x$ for $x \ll 1$, then we can write the condition for entanglement on the input state's squeezing parameter as $r > n$ for the initial state, and $r > n + \frac{1}{2}N_{\text{eff}}|r_{R}|^{2}$ for the output state.

We have found that it is not possible to achieve values of the reflection coefficient lower than $|r_{R}| \sim 0.08$ with a linear antenna. For the squeezing of the initial state around $r=1$, we need $|r_{R}| < 0.026$ in order for the output state to be entangled. An antenna in which the impedance grows linearly with the position is not sufficient, and for this we explore the stepwise antenna.

\begin{figure}[h]
{\includegraphics[width=0.5 \textwidth]{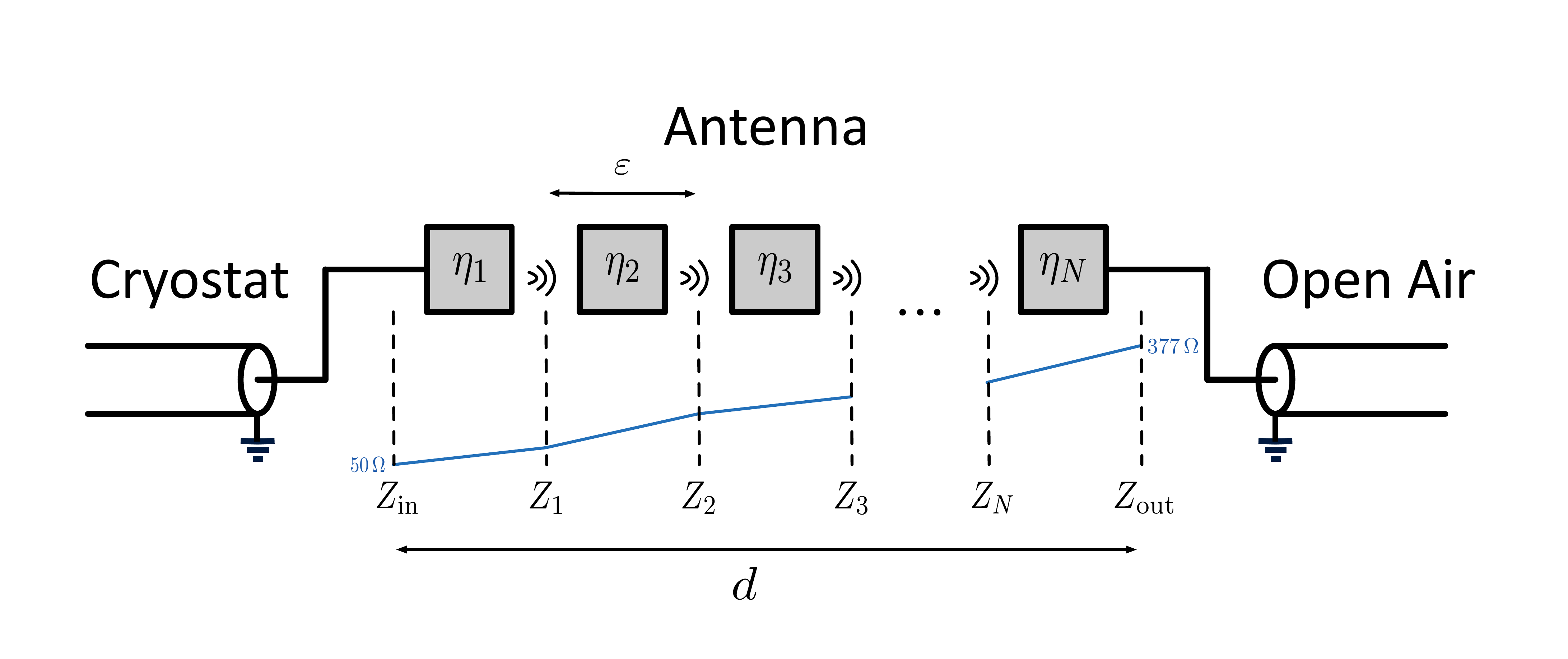}}
\caption{Quantum circuit design describing the connecting between the cryostat and the antenna, and between the antenna and the open air. Now, the antenna is divided in $N$ slices of length $\varepsilon$, inside which the impedance changes linearly, corresponding to a beam splitter with reflectivity $\eta_{i}=|r_{R}^{(i)}|^{2}$, setting $N+1$ scattering problems. Globally, we are able to implement a general function of the impedance.}
\label{fig3}
\end{figure}
 
\section{Stepwise antenna}\label{sec_4}
We propose a different approach to study the circuit: consider the division of the antenna in $N$ infinitesimally-small slices, regions in which the impedance changes linearly with the position, such that adding up all of these slices together yields an impedance that changes with the position following a different function. This setup can be seen in Fig.~\ref{fig3}. 

The difference with the previous approach is that now we have $N-1$ new parameters, the impedances of the intermediate slices, which we can use to optimize step by step the transfer of the quantum state in the antenna, together with the size of the TL. In this case, the impedance at a slice $m$ in the TL is given by
\begin{equation*}
Z(x) = \left( m+1-\frac{x}{\varepsilon}\right)Z_{m} + \left( \frac{x}{\varepsilon} -m \right)Z_{m+1}.
\end{equation*}
Then, the spatial component of the wavefunction for a given slice $m$ inside the antenna is given by
{\small\begin{eqnarray}\label{SL_solution_m}
 u_{m}(x) &&= [\varepsilon Z_{m}+(x-m\varepsilon)(Z_{m+1}-Z_{m})] \times \\
 \nonumber &&\left[ c_{1}^{(m)}J_{1}\left(k(x-m\varepsilon)+k\varepsilon\frac{Z_{m}}{Z_{m+1}-Z_{m}}\right)\right.\\
  \nonumber&&\left.+ c_{2}^{(m)}Y_{1}\left(k(x-m\varepsilon)+k\varepsilon\frac{Z_{m}}{Z_{m+1}-Z_{m}}\right)\right],
\end{eqnarray}}
where $\varepsilon=d/N$ indicates the size of each slice, for $x\in[\varepsilon m,\varepsilon (m+1)]$ and $m\in\{0, ..., N-1\}$. See that, for $N=1$, we recover the result of the linear antenna studied above. This system allows us to construct a transfer matrix for each of the $N$ scattering problems, such that the global transfer matrix will be the result of an ordered product of these $N$ matrices. In this problem, 
\begin{equation}
\begin{pmatrix} F \\ G \end{pmatrix} = T_{N}  \begin{pmatrix} c_{1}^{(N)} \\ c_{2}^{(N)} \end{pmatrix} = T_{N}...T_{0}\begin{pmatrix} A \\ B \end{pmatrix},
\end{equation}
and the global transfer matrix is $T = T_{N}T_{N-1}...T_{0}$. From this global transfer matrix, we can obtain the global scattering matrix, and make it unitary in the same way as for the linear antenna. This technique allows us to implement different continuous piecewise functions for the impedance, and provides more freedom in the optimization process. Eventually, the design of this circuit is oriented to optimize the resource that is shared between two parties. Thus, the optimization process will involve the minimization of the reflection coefficient $|r_{R}|$, in order to maximize the entanglement in the output TMSth state. 
\begin{figure}[t]
{\includegraphics[width=0.5 \textwidth]{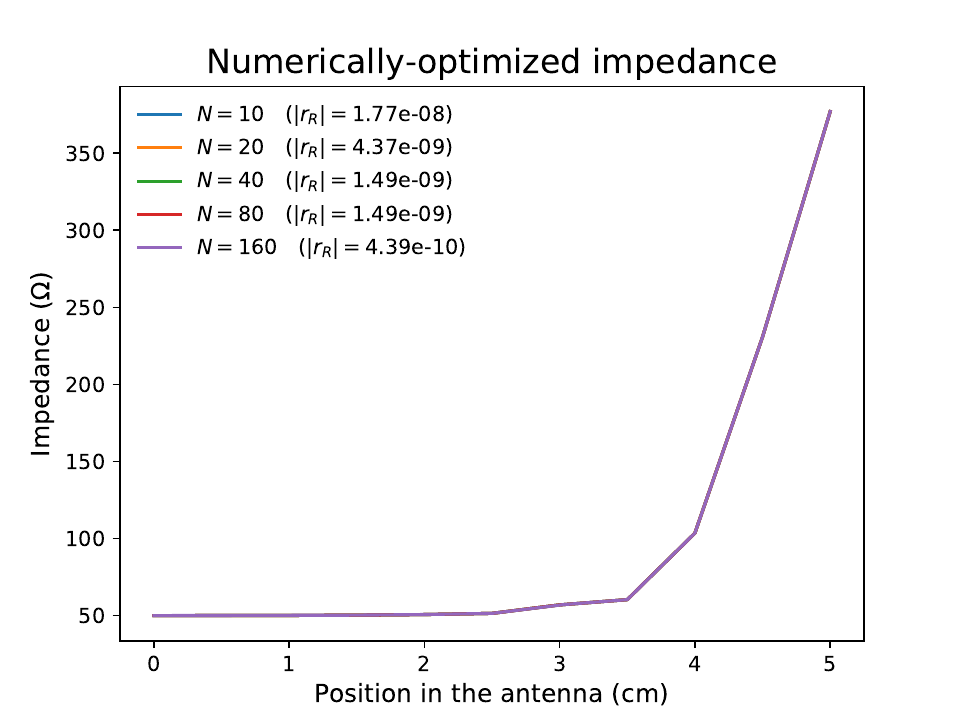}}
\caption{Numerically-optimized impedance curves against the position inside the antenna, of length $d = 5$ cm, for different values of the number of subdivisions in the antenna, $N$. Starting with the optimal solution for $N=10$, we compute the successive solutions through interpolation. For each one, the optimal value of the reflectivity $|r_{R}|$ is shown.}
\label{fig4}
\end{figure}
We are facing a global optimization problem that we will perform stepwise locally. Starting from random impedance arrays as initial guesses, we optimize the reflectivity with respect to the first impedance before $Z_{\text{out}} = 377 \, \Omega$, while keeping the rest of the impedances fixed. Once the optimal impedance value for the first point has been found, we update its value and optimize with respect to the previous point. We repeat the process for the whole impedance array until we reach the point before $Z_{\text{in}} = 50 \, \Omega$. Of course, $Z_{\text{in}}$ and $Z_{\text{out}}$ must remain fixed. As a criterion for the stability of the solutions, we consider that the optimization process has been successful when the difference between the reflectivities computed with the impedance solutions after two consecutive sweeps is smaller than $10^{-10}$. 

Even with just one subdivision ($N=2$), we are able to find small enough values of the reflection coefficient to have an entangled output state. In the solutions presented in Fig.~\ref{fig4}, we start from a small number of subdivisions ($N=10$) and optimize the reflectivity. We then interpolate the optimal impedance by doubling the number of slices, adding the average impedance value of every pair of points in the original array in between said points. This means splitting each slice in half, while keeping the same linear impedance function. Because of this, the new impedance array, for $N=20$, gives the same reflectivity as the optimal impedance array we found for $N=10$. Now, taking the interpolated array as the initial guess, we optimize the reflectivity for $N=20$, and continue in the same fashion until we reach $N=160$. In Fig.~\ref{fig4}, we can observe how the optimal impedance curves are shaped for different values of $N$, starting at $N=10$, and doubling it through interpolation, until $N=160$. For each value, we also give the value of the reflection coefficient that such an antenna could achieve. Notice that these values decrease as $N$ becomes larger, while the interpolation method leads to very small changes in the impedances, such that the curves cannot be distinguished.

\begin{figure}[t]
{\includegraphics[width=0.5 \textwidth]{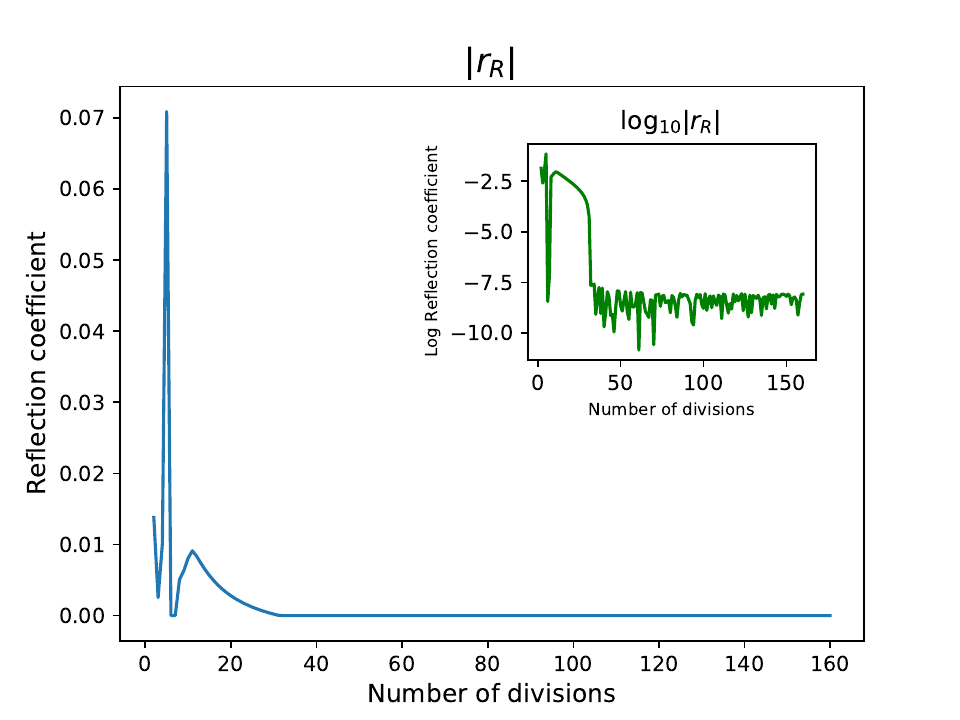}}
\caption{Reflection coefficient computed with the impedance function proposed in Eq.~\ref{ansatz}, represented for different values of the number of subdivisions inside the antenna, for an antenna of size $d=5$ cm. As an inset, we show $\log_{10}|r_{R}|$ to illustrate how this impedance function reduces the reflection coefficient down to $10^{-8}$.}
\label{fig5}
\end{figure}

To speed up the optimization process and try to better recognize the behavior of the optimal impedance, we propose an ansatz to describe it,
\begin{equation}\label{ansatz}
{\small Z(x) = Z_{\text{in}} + \alpha \left[ e^{\left(\frac{x}{d}\right)^{\beta}\log\left( 1+\frac{Z_{\text{out}}-Z_{\text{in}}}{\alpha}\right)} - 1\right]}
\end{equation}
where $d$ is the size of the antenna, $x$ indicates the position inside it, and $\alpha$, $\beta$ are free parameters that we can optimize. This functional ansatz is an inspiration on the qualitative behavior of the curves in Fig.~\ref{fig4}, and not an actual fit of the numerical data. Our goal is to rewrite $N-1$ local numerical optimization problems as a global optimization problem with just two parameters, $\alpha$ and $\beta$, in order to improve convergence and stability of the solutions. Notice that the results we will find using this function will differ from those obtained with numerical optimization. In fact, since this is only an approximation of the optimal solution, the reflectivities we compute with this exponential impedance will be larger than those we can obtain with numerical optimization. We have found the optimal values to be $\alpha \sim 10.31$ with $\beta \sim 0.69$, for $d = 5$ cm. See that, for $\alpha \rightarrow \infty$, we recover the linear antenna. 

This function approximates the behavior of the optimal impedance on the antenna, but the values of the reflection coefficient we obtain with it are not small enough. However, these improve as we increase $N$, as can be seen in Fig.~\ref{fig5}, oscillating around $|r_{R}| \sim 10^{-8}$ for $N$ approaching 160. We observe that minimal values of $|r_{R}|$ are achieved for $N>30$, which must represent a regime where $\varepsilon = d/N \ll \lambda$, approaching the continuum limit. This promising result suggests that we could employ the same treatment of the antenna as we did for a linear impedance, but solving the Sturm-Liouville problem with the impedance given by Eq.~\ref{ansatz}, in the limit $N=1$.

\begin{figure}[t]
{\includegraphics[width=0.5 \textwidth]{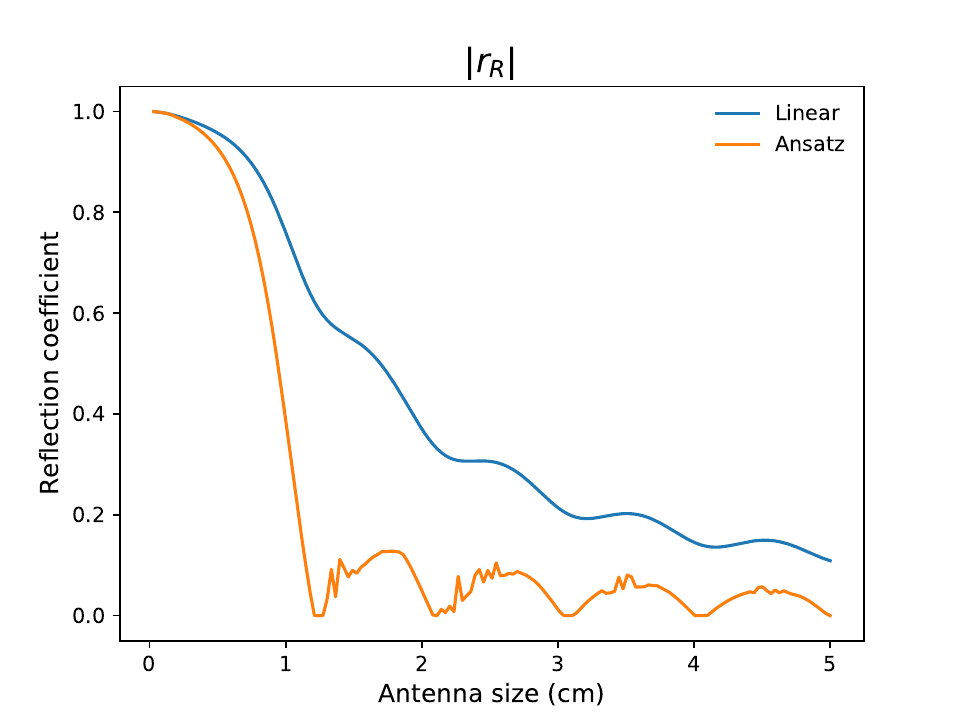}}
\caption{Reflection coefficient against the size of the antenna. The reflection coefficient is computed with a linear function of the impedance (blue) and the impedance ansatz proposed in Eq.~\ref{ansatz} (orange), for $N=160$. Notice that the reflectivity decreases further as we increase the size of the antenna, continuously for a linear impedance, and jumping between minimum values for the ansatz.}
\label{fig6}
\end{figure}

Taking $N=160$, we represent the reflection coefficient versus the antenna size in Fig.~\ref{fig6}. In blue, we observe the reflectivity of the antenna with a linear impedance function, and in orange we see the result of the reflectivity corresponding to the impedance function proposed in Eq.~\ref{ansatz}, with optimized parameters. We observe that minimal values of $|r_{R}|$ are achieved for particular values of the antenna size, which approximately coincide with multiples of half the wavelength inside the antenna. Also, we observe that, in order to find optimal values of the reflectivity, we require $d>\lambda/2$.

\begin{figure}[t]
{\includegraphics[width=0.5 \textwidth]{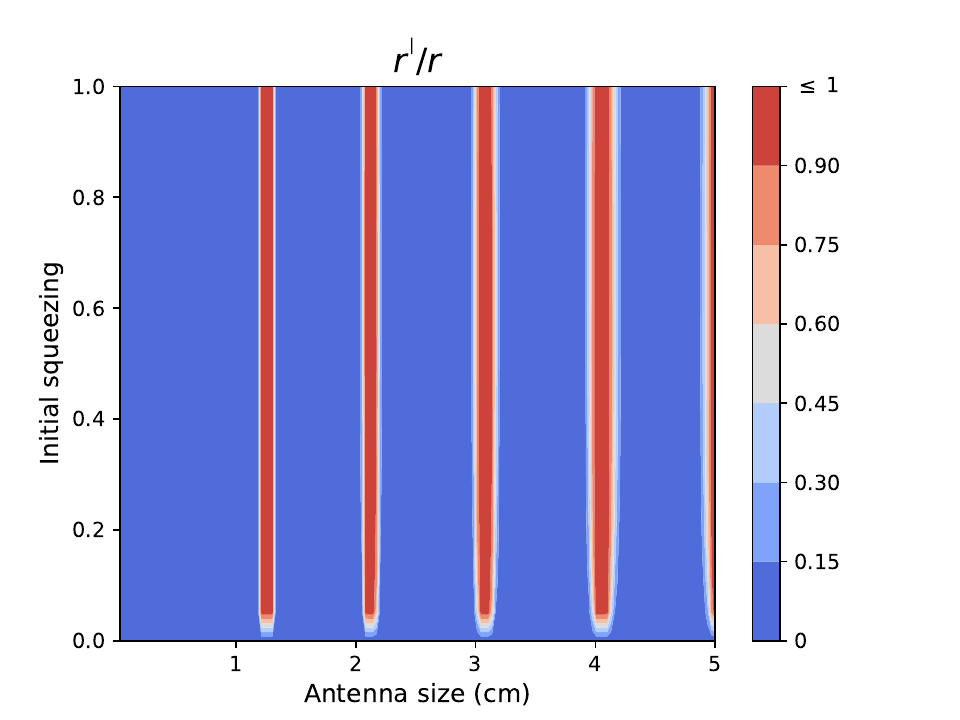}}
\caption{Ratio between squeezing of the output state $r'$ and squeezing of the input state $r$, represented over $r$ and over the size of the antenna, for an antenna with $N=160$ subdivisions. This shows that at least $90$~\% of the initial squeezing can be recovered with an antenna of size equal to the multiples of half a wavelength, with initial squeezing $r>0$.}
\label{fig7}
\end{figure}

Finally, we investigate the squeezing of the output state, in terms of the initial squeezing and the size of the antenna. In Fig.~\ref{fig7} we represent the quotient between squeezing parameters of output and input states, showing that it is possible to preserve squeezing in the multiples of the half-wavelength of the signal, the same spots we found in Fig.~\ref{fig6} for the size of the antenna, where the reflectivity is minimal.

In order to illustrate the sensitivity of the reflection coefficient to the shape of the antenna, we introduce errors to the numerically-optimized impedance as a random value taken from a normal distribution where the variance is a percentage of the value of the function at each point. With this modified impedance we compute the reflection coefficient, and then calculate the ratio between the negativity of the output state and the negativity of the input state, $\mathcal{N}_{\text{out}}/\mathcal{N}_{\text{in}}$. This study indicates a limit on manufacturing errors oriented towards the fabrication of such a device. 

\begin{figure}[t]
{\includegraphics[width=0.5 \textwidth]{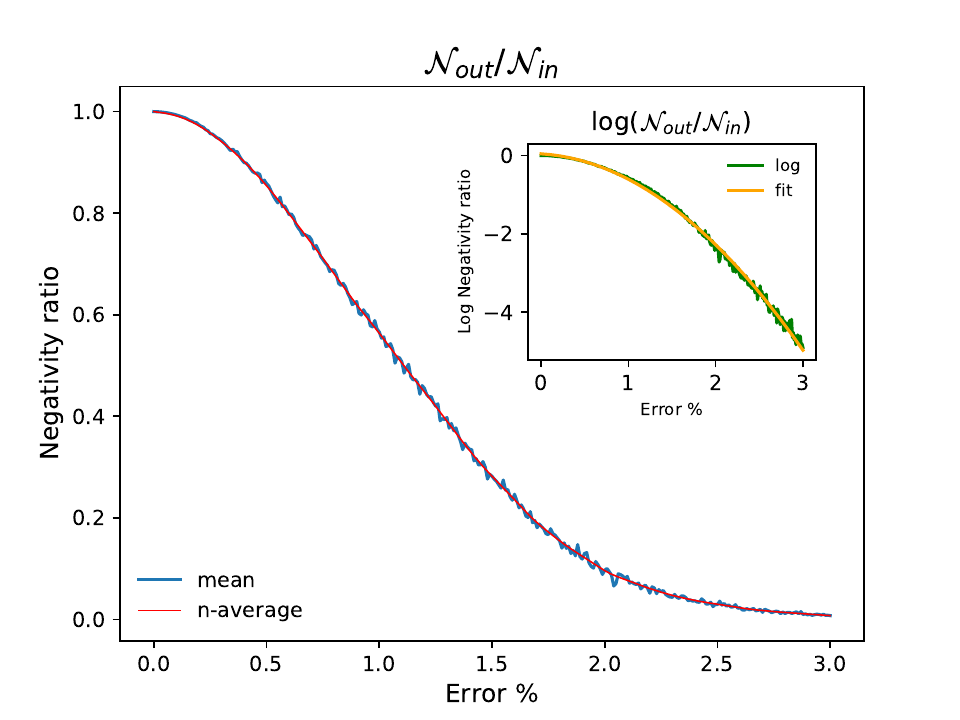}}
\caption{Ratio between negativity of the output state $\mathcal{N}_{\text{out}}$ and negativity of the input state $\mathcal{N}_{\text{in}}$, averaged over many iterations in which the impedance function is modified with a random error proportional to a percentage of the value of the impedance at each point. In blue, we represent the mean value of the negativity ratio over different error percentages, and in red we display the smoothing of the mean by applying a n-average technique (see Appendix~\ref{app2}), for $n=50$. In green, the inset shows the logarithm of the ratio between negativities, and in orange we show a quadratic fit of the logarithm of the negativity ratio. The latter corresponds to a function $ax^{2}+bx+c$ with $a\sim -0.51$, $b\sim -0.14$, and $c\sim 0.04$, and with variance $\sim 0.01$.}
\label{fig8}
\end{figure}

In Fig.~\ref{fig8}, we represent the average ratio of negativities for different values of the error percentage (blue), and we observe that it decreases as we increase the error, for an initial squeezing $r=1$, $N=160$ subdivisions and antenna size $d = 5$ cm. As an inset, we represent the logarithm of the ratio of negativities (green), which we fit by a quadratic function (orange), as the function seems to follow a gaussian. In red, we represent the $n$-average (see Appendix~B) of the mean negativity ratio, a function towards which the mean should tend to in an infinite-trial scenario. Here, each error percentage step is averaged $10^{3}$ times, and we have taken $n=50$ for the $n$-average. 

The results show that the negativity ratio goes to zero for errors over 3\% of the impedance values, and from the quadratic fit of the logarithm, we can extract a function $ax^{2}+bx+c$ with $a\sim -0.51$, $b\sim -0.14$, and $c\sim 0.04$, and with variance $\sim 0.01$.

\section{Antenna Design}
In this article, we have proposed an antenna based on a coplanar waveguide, and the characteristics of this waveguide will depend on the impedance we want to implement. Consider a coplanar waveguide, whose central conducting plate has a width of $2a$ and a height much smaller than the total depth of the film, and in which the distance between the middle of the conducting plate and the start of the grounded plates is $b$. By defining $\rho=a/b$, we can write the density of inductance and the density of capacitance for such a waveguide as~\cite{Goppl2008,Clem2013}
\begin{eqnarray}
\label{CPW_L} l &=& \frac{\mu_{0}}{4}\frac{K\left(\sqrt{1-\rho^{2}}\right)}{K(\rho)}, \\ 
\label{CPW_C} c &=& 4\varepsilon_{0}\varepsilon_{\text{eff}}\frac{K(\rho)}{K\left(\sqrt{1-\rho^{2}}\right)},
\end{eqnarray}
where $\mu_{0}$ and $\varepsilon_{0}$ are the magnetic permeability and the electric permittivity of the vacuum, respectively, and $\varepsilon_{\text{eff}}$ is the effective dielectric constant; it is a function of the geometry of the waveguide, but also of the permittivities of the substrate and the oxide layers. Here, we have defined $K(y)$ as the complete elliptic integral of the first kind with modulus $y$, such that
\begin{equation}
K(y) = \int_{0}^{\pi/2} \text{d}\theta \frac{1}{\sqrt{1-y^{2}\sin^{2}\theta}}.
\end{equation}
From Eqs.~\ref{CPW_L},~\ref{CPW_C}, the characteristic impedance of the waveguide is straightforwardly obtained,
\begin{equation}
Z = \bar{z}\frac{K\left(\sqrt{1-\rho^{2}}\right)}{K(\rho)},
\end{equation}
with $\bar{z} = \frac{1}{4}\sqrt{\frac{\mu_{0}}{\varepsilon_{0}\varepsilon_{\text{eff}}}} = 10\pi$. For the cryostat impedance $Z_{\text{in}} = 50\,\Omega$, this requires that $\rho_{\text{in}}\approx 0.32$, and for the impedance of open air, $Z_{\text{out}} = 377\,\Omega$, $\rho_{\text{out}}\approx 2.60 \cdot 10^{-8}$.

In order to implement the kind of antenna we have proposed here, a coplanar waveguide has to be designed with a varying ratio $\rho$. One way to do this is to solve the above equation, for each value of $Z$. The dependence of $\rho$ on the position inside the antenna could be inferred by substituting the values of $Z$ by those given in the ansatz we propose in Eq.~\ref{ansatz}. Alternatively, we could directly propose an ansatz for $\rho$, targeting a function of the position in the antenna that leads to an easier technological design. Similar to what we did previously, certain parameters in this ansatz should be left open in order to optimize over them to obtain the ideal impedance. A simple example would be to consider
\begin{equation}
\rho(x) = \rho_{\text{in}} + \alpha \left[ e^{\left(\frac{x}{d}\right)^{\beta}\log\left( 1+\frac{\rho_{\text{out}}-\rho_{\text{in}}}{\alpha}\right)} - 1\right].
\end{equation}
Usual values of $a$ and $b$ are 5 $\mu$m and 7 $\mu$m, respectively. Fixing the value of $a_{\text{in}}$ to 5 $\mu$m, we would need $b_{\text{in}}=15.63$ $\mu$m in order to obtain $\rho_{\text{in}}$. To get $\rho_{\text{out}}$ at the termination of the antenna, we could for example set $a_{\text{out}} = 10$ nm and $b_{\text{out}} = 38.46$ cm. In principle, this may be achieved, given that the electron-beam lithography can achieve a precision below 10 nm for the fabrication of coplanar waveguides. However, the London depth of the material will impose a lower bound on the value of $a$ we would ideally want to set. Different realizations of such a device could be based on carbon-nanotube ink deposits on the gap of the coplanar waveguide, as described in Ref.~\cite{DePaolis2011}, or on coplanar waveguides with width-varying superconducting plate, studied in Ref.~\cite{Shamaileh2019}.


Throughout this article we have assumed that the antenna is implemented in a superconducting TL, meaning that the temperature inside it is in the range of mK (or at least below $4$ K). However, this would be very difficult to implement, since the end of this line is connected to the open air, whose temperature is 300 K. In order to maintain a low temperature in the antenna, with a constant propagation velocity of $v_{\text{in}}=c/3$, and still be able to connect it to the open air, we could study the addition of a subsequent waveguide. It would have the impedance of open air, 377 $\Omega$, while presenting a temperature gradient, as well as a velocity gradient, from $c/3$ to $c$. 

We consider modeling absorption losses due to loss of superconductivity in a transmission line of length $L$ that connects the antenna, at cryogenic temperatures, with the open air at $300$~K, by an infinite array of beamsplitters. Each beam splitter has a reflectivity $\eta_{i}$ that represent absorption probability, and incorporates thermal noise at a given temperature $T_{i}$ inside the TL, characterized by a number of thermal photons $n(T_{i})$. This infinite array of beamsplitters can be represented by a single effective beam splitter, with reflectivity $\eta_{\text{eff}} = 1-e^{\int_{0}^{L}dx\mu(x)}$, where $\mu(x)$ is the reflectivity density as a function of the positions inside the TL. The effective number of thermal photons that this beam splitter incorporates to the system (see Appendix~\ref{app3}) is
\begin{equation}
n_{\text{eff}} = \frac{\int_{0}^{L} dx \mu(x)n(x) e^{-\int_{x}^{L}dx' \mu(x')}}{1-e^{-\int_{0}^{L} dx \mu(x)}}.
\end{equation}
This expression is general and can be applied to any case in which we know the profile of temperatures. Let us now choose a simple but useful profile which allows us to find a closed expression. Indeed, if we consider that the TL can be kept at temperatures below the critical one for a length $L_{0} < L$, then we can choose
\begin{eqnarray*}
\nonumber n(x) &=& n(T_{\text{in}}) + [n(T_{\text{out}})-n(T_{\text{in}})]\theta(x-L_{0}), \\
\mu(x) &=& \mu_{\text{in}} + (\mu_{\text{out}}-\mu_{\text{in}})\theta(x-L_{0}),
\end{eqnarray*}
where $\mu_{\text{in}}$ describes absorption losses at cryogenic temperatures and $\mu_{\text{out}}$ describes absorption losses of the material at room temperature. Then, the effective number of thermal photons becomes
\begin{eqnarray}
\nonumber n_{\text{eff}} &=& n(T_{\text{in}}) \frac{e^{-\mu_{\text{out}}(L-L_{0})}(1-e^{-\mu_{\text{in}}L_{0}})}{1-e^{-\mu_{\text{in}}L_{0}}e^{-\mu_{\text{out}}(L-L_{0})}} \\
&+& n(T_{\text{out}})\frac{1-e^{-\mu_{\text{out}}(L-L_{0})}}{1-e^{-\mu_{\text{in}}L_{0}}e^{-\mu_{\text{out}}(L-L_{0})}}.
\end{eqnarray}
Notice that, when $L_{0}=0$, then $n_{\text{eff}} = n(T_{\text{out}})$ and, when $L_{0}=L$, then $n_{\text{eff}} = n(T_{\text{in}})$, as expected. Consequently, for $T_{\text{out}}=300$~K and $\omega/2\pi = 5$~GHz, and by using the Bose-Einstein distribution, we obtain that $n(T_{\text{out}}) \approx N_{\text{eff}}\sim1250$, which is considered as the input thermal noise into the antenna. The number of thermal photons at cryogenic temperatures is $n\sim8\cdot10^{-3}$, corresponding to $T_{\text{in}}=50$~mK and the same frequency. Given that $n/N_{\text{eff}}\sim10^{-6}$, we have
\begin{equation}
\frac{n_{\text{eff}}}{N_{\text{eff}}} \approx \frac{1-e^{-\mu_{\text{out}}(L-L_{0})}}{1-e^{-\mu_{\text{in}}L_{0}}e^{-\mu_{\text{out}}(L-L_{0})}} \leq 1,
\end{equation}
since $e^{-\mu_{\text{in}}L_{0}}\leq 1$. This implies that, considering this approach, the effect of thermal noise in the antenna is reduced when compared with respect to the study we present here. The reason is that before, we were considering the thermal state as the incoming state of the antenna from the right, while it is now substantially reduced since part of the thermal photons are also absorbed in the cryostat before arriving at the antenna. Therefore, the introduction of these losses is a tradeoff between the effect of the effective beam splitter on the entanglement, and the improvement on the performance of the antenna due to the lower number of photons corresponding to the effective thermal state. Of course, these effects will substantially depend on the exact profile of temperatures along the TL. When this is obtained, one should repeat the optimization procedure for the impedance and then add the effective beam splitter after the antenna to take into account the entanglement degradation. 

Although it is a crucial part of classical microwave communication, amplification of signals is not relevant in this setup. Consider a cryogenic HEMT amplifier, currently used in quantum microwave experiments, which produces large gains in a relatively large frequency spectrum, but also introduces a significant amount of noise. Thermal noise added by commercial HEMTs is counted in the range $n\sim10-100$ photons in the considered frequency regime~\cite{DiCandia2015}. Since the amplification is applied to the modes individually, it results ideal to enhance classical signals, but it cannot increase quantum correlations. To sum up, the goal of this work is to preserve quantum correlations in open air and traditional amplification does not provide an advantage in this objective, on the contrary, it could lead to entanglement degradation due to the introduction of thermal noise. 

\section{Conclusions \& Perspectives}
In this article we have studied the optimization of a quantum circuit for an open-air microwave quantum communication protocol, in which entangled states, which are generated in a cryostat, are sent into the air, maximizing the entanglement of the output state in the presence of thermal noise. This circuit consists of a waveguide that transports the state out of the cryostat, and a waveguide representing the transmission of that state in open air, both connected by an antenna that realizes a smooth impedance matching between the two environments, maximizing the transmission of energy. 

Knowing that previous studies using similar architectures had failed to detect entanglement in open air, we have investigated the simplest case of quantum antenna, a finite inhomogeneous transmission line with an impedance that changes linearly with the position. Studying the transmission of two-mode squeezed thermal states, we have found that such a device cannot preserve entanglement due to insufficiently low reflectivity. 

For that, we have proposed a stepwise antenna that introduces $N$ subdivisions, inside which the local impedance grows linearly, but globally can implement a more general function. Numerically, we were able to find a shape of the impedance which minimized the reflectivity of the antenna enough to preserve entanglement on the output state in the presence of thermal noise. Inspired by this shape, we proposed an exponential function to describe the optimal impedance. This shape leads to a reflectivity that decreases with higher $N$, as if taking the continuum limit, and that improves significantly with respect to a linear shape. In fact, the reflectivities are low enough as to being able to preserve over 90\% of the squeezing of the initial state. However, it cannot improve the numerical results previously obtained. A further simulation confirms that errors over 3\% of the values of the impedance function result in a destruction of the entanglement, exemplifying how easily an entanglement distribution protocol can be truncated by the use of a simple antenna. 

This works will impact the fields of quantum illumination and quantum sensing, with particular emphasis on the quantum radar, as well as any quantum communication protocol dependent on entanglement distribution in the microwave regime. 

\acknowledgements
The authors are grateful to Kirill G. Fedorov, Frank Deppe, Benjamin Huard, and Shabir Barzanjeh for useful comments and discussions. 
\\
The authors acknowledge financial support from Basque Government QUANTEK project from ELKARTEK program (KK-2021/00070), Spanish Ram\'on y Cajal Grant RYC-2020-030503-I and the project grant PID2021-125823NA-I00 funded by MCIN/AEI/10.13039/501100011033 and by ``ERDF A way of making Europe'' and ``ERDF Invest in your Future'', as well as from QMiCS (820505) and OpenSuperQ (820363) of the EU Flagship on Quantum Technologies, and the EU FET-Open projects Quromorphic (828826) and EPIQUS (899368).

\bibliographystyle{quantum}

\onecolumn
\appendix

\section{Sturm-Lioville problem for the wavefunction inside the antenna}\label{app1}
In this appendix, we solve the Sturm-Liouville problem for the spatial part of the wavefunction, in the case of an antenna whose impedance is a linear function. The equation we want to solve is
\begin{equation}
u''(x)-\frac{Z'(x)}{Z(x)}u'(x) + k^{2}u(x) = 0,
\end{equation}
and first we will multiply by $(Z(x)/Z'(x))^{2}$, resulting in 
\begin{equation}
\left(\frac{Z(x)}{Z'(x)}\right)^{2}u''(x)-\frac{Z(x)}{Z'(x)}u'(x) + k^{2}\left(\frac{Z(x)}{Z'(x)}\right)^{2}u(x) = 0.
\end{equation}
For a linear impedance $Z(x) = \left(1-\frac{x}{d}\right)Z_{\text{in}} + \frac{x}{d} Z_{\text{out}}$, we have 
\begin{equation}
\frac{Z(x)}{Z'(x)} = x + d\frac{Z_{\text{in}}}{Z_{\text{out}}-Z_{\text{in}}}.
\end{equation}
Let us define $y = k Z(x)/Z'(x)$, such that $u'(x) = k u'(y)$, $u''(x) = k^{2} u''(y)$. With this, we can rewrite
\begin{equation}
y^{2}u''(y)-y u'(y) + y^{2}u(y) = 0.
\end{equation}
If we introduce $y\varphi(y)=u(y)$, with $u'(y)=\varphi(y)+y\varphi'(y)$ and $u''(y)=2\varphi'(y)+y\varphi''(y)$, we find 
\begin{equation}
y^{2}\varphi''(y)+y \varphi'(y) + (y^{2}-1)\varphi(y) = 0,
\end{equation}
which is the first order Bessel's differential equation. The solution to this equation is
\begin{equation}
\varphi(y) = b_{1} J_{1}(y) + b_{2} Y_{1}(y),
\end{equation}
and if we undo all the variable changes, we find
\begin{equation}
u(x) = k\left( x + d\frac{Z_{\text{in}}}{Z_{\text{out}}-Z_{\text{in}}}\right) \bigg[ b_{1} J_{1}\left( k x + k d\frac{Z_{\text{in}}}{Z_{\text{out}}-Z_{\text{in}}}\right) + b_{2} Y_{1}\left( k x + k d\frac{Z_{\text{in}}}{Z_{\text{out}}-Z_{\text{in}}}\right) \bigg].
\end{equation}
To obtain the result shown in Eq.~\ref{SL_solution}, we need to redefine $b_{i} k/(Z_{\text{out}}-Z_{\text{in}}) = c_{i}$, with $i=\{1,2\}$.

This derivation is equal to the one that leads to Eq.~\ref{SL_solution_m}, for the spatial component of the wavefunction inside slice $m$ in the stepwise antenna. In this case, the only difference is that we have an impedance 
\begin{equation}
Z(x) = \left( m+1-\frac{x}{\varepsilon}\right)Z_{m} + \left( \frac{x}{\varepsilon} -m \right)Z_{m+1},
\end{equation}
and then, parameter $y$ will be defined as
\begin{equation}
y = k\frac{Z(x)}{Z'(x)} = k(x-m\varepsilon) + k\varepsilon\frac{Z_{m}}{Z_{m+1}-Z_{m}}.
\end{equation}
Then, the arbitrary parameters of the solution need to be redefined as $b_{i}^{(m)} k/(Z_{m+1}-Z_{m}) = c_{i}^{(m)}$, for $i=\{1,2\}$.

\section{$n$-average technique for function smoothing}\label{app2}
Take a discrete function $f_{0}$, evaluated over a grid of points labelled by $x_{k}$, for $k\in[0,L]$. This function results from an average over many trials, given that it has a stochastic component based on a normal distribution. The function still presents traces of stochastic behavior, since the number of trials we can perform is finite. Our goal is to find the value towards which the infinite average of the function tends. For that, we propose the computation of the average of the function on a given point, such that 
\begin{equation}
f_{1}(x_{k}) = \frac{f_{0}(x_{k+1})+2f_{0}(x_{k})+f_{0}(x_{k-1})}{4},
\end{equation}
where $f_{1}$ is the 1-averaged function. Then, the $n$-averaged function is
\begin{equation}
f_{n}(x_{k}) = 2^{-2n} \sum_{m=0}^{2n} \begin{pmatrix} 2n \\ m \end{pmatrix} f_{0}(x_{k+n-m})\theta(k+n-m),
\end{equation}
with $\theta(0)=1$ and $\begin{pmatrix} 2n \\ m \end{pmatrix}=\frac{2n!}{m!(2n-m!)}$. Here, $n$ indicates the number of times the average has been performed, $k$ represents a point where the function is evaluated, and $m$ is a dummy index of the sum that goes through all the values that contributed to the $n$-average of the function at a point $x_{k}$. If $n > k$, then $m\in[0,n+k]$, and if $n < k$, $m\in[0,2n]$. From our definition of average we have taken $f_{n}(x_{0})=...=f_{1}(x_{0})=f_{0}(x_{0})$ and $f_{n}(x_{L})=...=f_{1}(x_{L})=f_{0}(x_{L})$. 

The largest binomial coefficient, $\begin{pmatrix} n \\ m \end{pmatrix}$, occurs at $m=n/2$, and then the largest contribution to the weighted sum that represents the $n$-average is
\begin{equation}
f_{n}(x_{k}) \approx 2^{-2n} \begin{pmatrix} 2n \\ n \end{pmatrix} f_{0}(x_{k}).
\end{equation}
This process exemplifies a discrete, binomial convolution, which in the continuum limit becomes a gaussian convolution. 

\section{Absorption losses by an infinite beam splitter array}\label{app3}
Consider a beam splitter with reflectivity $\eta_{i}$ that incorporates thermal noise to a signal as the reflected contribution. The output mode of an array of N beamsplitters of this kind is given by
\begin{equation}
a_{N} = a_{1}\prod_{i=1}^{N-1}\sqrt{1-\eta_{i}} + \sum_{k=1}^{N-1}h_{k}^{\text{in}} \sqrt{\eta_{k}}\prod_{i=k+1}^{N-1}\sqrt{1-\eta_{i}}
\end{equation}
for an input signal mode $a_{1}$, where the number of thermal photons incorporated by beam splitter $k$ is given by $n_{k}=\langle h_{k}^{\text{in}\dagger}h_{k}^{\text{in}}\rangle$. We aim at representing the action of this array of beamsplitters by a single beam splitter with effective reflectivity and effective number of thermal photons. In this expression, we can identify effective reflection and transmission coefficients,
\begin{equation}
a_{N} = a_{1}\sqrt{1-\eta_{\text{eff}}} + h_{\text{eff}}^{\text{in}} \sqrt{\eta_{\text{eff}}}.
\end{equation}
Consider the reflectivity of a beam splitter as $\eta_{i}=\mu L/N$, where $\mu$ is the reflectivity per unit length. For very large N, assume $L/N = \Delta x$. Then, we could write $\eta_{i}=\mu_{i} \Delta x$, and then the effective reflectivity is simplified by
\begin{equation}
\log(1-\eta_{\text{eff}}) = \sum_{i=1}^{N-1}\log(1-\eta_{i}) = \sum_{i=1}^{N-1}\log(1-\mu_{i}\Delta x).
\end{equation}
For $\Delta x \ll 1$, we can expand this as $\log(1-\mu_{i}\Delta x) \approx -\mu_{i}\Delta x$, and taking the continuum limit, 
\begin{equation}
-\sum_{i=1}^{N-1}\mu_{i}\Delta x \longrightarrow -\int_{0}^{L} dx \mu(x).
\end{equation}
Then, we write $\eta_{\text{eff}} = 1-e^{-\int_{0}^{L} dx \mu(x)}$. Let us now compute the effective number of thermal photons, 
\begin{equation}
\eta_{\text{eff}}n_{\text{eff}} = \langle (\sqrt{\eta_{\text{eff}}}h_{\text{eff}})^{\dagger}(\sqrt{\eta_{\text{eff}}}h_{\text{eff}})\rangle = \sum_{k=1}^{N-1}\eta_{k} n_{k} \left[\prod_{i=k+1}^{N-1}(1-\eta_{i})\right],
\end{equation}
which can be expressed as
\begin{equation}
\eta_{\text{eff}}n_{\text{eff}} = \sum_{k=1}^{N-1}\mu_{k}\Delta x \, n_{k} e^{-\int_{x}^{L}dx' \mu(x')} = \int_{0}^{L} dx \mu(x)n(x) e^{-\int_{x}^{L}dx' \mu(x')}.
\end{equation}
Then, the effective number of thermal photons is
\begin{equation}
n_{\text{eff}} = \frac{\int_{0}^{L} dx \mu(x)n(x) e^{-\int_{x}^{L}dx' \mu(x')}}{1-e^{-\int_{0}^{L} dx \mu(x)}}.
\end{equation}

\end{document}